# Dynamical patterns of phase transformations from self-trapping of quantum excitons

Tianyou Yi[a], Natasha Kirova[b,d,*], Serguei Brazovskii[c,d]

[a] *South University of Science and Technology of China, Shenzhen, Guangdong 518055, China*
[b] *CNRS, LPS, URM 8626, Univeristé Paris-sud, Orsay, 91405, France*
[c] *CNRS, LPTMS, URM 8502, Univeristé Paris-sud, Orsay, 91405, France*
[d] *International Institute of Physics, UFRN, 59078-400 Natal, RN, Brazil*

**Abstract**

Phase transitions induced by short optical pulses is a new mainstream in studies of cooperative electronic states. Its special realization in systems with neutral-ionic transformations stands out in a way that the optical pumping goes to excitons rather than to electronic bands. We present a semi-phenomenological modelling of spacio-temporal effects applicable to any system where the optical excitons are coupled to a symmetry breaking order parameter. In our scenario, after a short initial pulse of photons, a quasi-condensate of excitons appears as a macroscopic quantum state which then evolves interacting with other degrees of freedom prone to instability. This coupling leads to self-trapping of excitons; that locally enhances their density which can surpass a critical value to trigger the phase transformation, even if the mean density is below the required threshold. The system is stratified in domains which evolve through dynamical phase transitions and may persist even after the initiating excitons have recombined. We recover dynamic interplays of fields such as the excitons' wave function, electronic charge transfer and polarization, lattice dimerization.

## 1. Introduction

Nowadays, transformations among cooperative electronic states can be performed by short optical pulses [1-4]. By these pump induced phase transitions (PIPT), various symmetry broken ground states are being tested: crystallization of electrons (charge order with ferroelectricity) or of electron-hole (e-h) pairs (charge/spin density waves – CDW,SDW), the superconductivity, Peierls and Mott insulators.

The theory of PIPTs faces very high challenges when started *ab initio* at the microscopic level ([5] for review, [6]). But over longer time scales the evolution is governed by collective variables like an order parameter and lattice deformations. Effectiveness of such a phenomenological approach has been proven by detailed modelling of coherent dynamics of a macroscopic electronic order through destruction and recovering of the CDW state. That allowed to recover such effects as dynamic symmetry breaking, stratification in domains and subsequent collapses of their walls, all in detailed accordance with the experiment [7]. Another example was a modelling [8] of the recently discovered [9] switching to a truly stable hidden state of a polaronic Mott insulator in 1T-TaS$_2$.

---

[*] Corresponding author. Tel.: +33-1-60 95; e-mail: kirova@lps.u-psud.fr.

In most experiments on PIPTs, the electrons are excited to a high energy with their subsequent very fast cooling down to a quasi-equilibrium e-h population which provokes events of a subsequent evolution. A new situation can take place if the photons are tuned in resonance with excitations, whatever is there origin – intra-molecular excitons (IME) or bound electron-hole pairs – charge transfer excitons (CTE). With so high concentration (up to 10% in PIPT experiments or 1% in our modelling below), the quasi-condensate of excitons should appear as a macroscopic quantum state which then evolves interacting with other degrees of freedom. Consequently a theory [10] can better bridge between the classical macroscopic and the quantum microscopic regimes.

The common case of pumping to unbound electrons and holes is not excluded from this scenario, provided the early cooling leads first to formation of excitons; this is what is known to take place in light emitting polymers [11] or in conventional semiconducting lasers operating at low temperature (T).

Particularly, we are interested in effects of excitons' self-trapping [12-14], akin to self-focusing in the nonlinear optics [15] or to formation of polarons from electrons. The locally enhanced density of excitons can surpass a critical value to trigger a phase transformation in another coupled degree of freedom, even if the mean density $n$ is below the required threshold.

In this article we shall demonstrate this principle on an experimentally elaborated example of optically provoked transformation between neutral N and ionic I phases in organic crystals like TTF-CA (see [16] and refs. therein). We shall build a minimal phenomenological model for IMEs which interaction with the inter-molecular charge transfer $\rho$ leads finally to symmetry breaking lattice dimerizations and hence to the ferroelectric polarization. The numerical modeling will be performed showing formation, evolution, abrupt transitions, and sometimes the decay of sharply localized domains of the embedded new phase. As an application, we can already target earlier experiments [17,18] with the pumping into the high energy mode of the TTF molecule. A more specialized model will be necessary in the future to describe pumping to the lower energy inter-molecular CTE [19-21] where the internal structure of the exciton starts to play an important role.

**2. Phenomenological approach to the N-I first order phase transition**

*2.1.  The model free energy.*

The NI transition takes place in a bi-molecular donor-acceptor chain (e.g. TTF-CA, see refs. in [16]) which shows a variable charge transfer (TTF$^{\rho}$-CA$^{-\rho}$) between the lower $\rho_N$ in the quasi-neutral (N) high T phase and the higher $\rho_I$ in the low T ionic (I) phase.

The 1$^{st}$ order transition in $\rho$ alone would go without a symmetry breaking and could be described by a generic double-well curve of Fig.1 for the free energy $W(\rho)$ with two minima at $\rho_N$ and $\rho_I$. The critical increase of the dielectric constant [16] tells us that, in spite of the N and

I minima being quite distant (e.g. $\rho_N=0.32$ and $\rho_I=0.52$ at the phase coexistence), the separating barrier is small.

In TTF-CA and the family, the situation is more interesting because another degree of freedom – alternating molecular displacements $h$ are involved: the I phase is accompanied by the lattice dimerization, so there is a symmetry breaking and the transition could have been of the 2nd order which is not the case nevertheless – the jump in $h$ is concomitant with the one in $\rho$. (This observation concerns the transition under temperature, while may change under pressure [22].) But after the pumping, as we shall show, a sharp dynamical transition should be observed.

There are several ways to phenomenologically guess the appropriate function for the ground state energy [23-25]. Ours is different in two respects. First, we minimize the suggestions to the level that the double shape appears only from interactions of $\rho$ and $h$ without enforcing it to be already a property of $\rho$ alone. Most importantly, we introduce the field of the multi-exciton coherent state as an integral part of the energy functional. Still we shall not take into account particular features of CTEs keeping the form as generic as possible for applications in other possible situations in other materials.

The system will be described by the free energy as a functional of $q(x,t)=\rho-\rho_n$, lattice deformations $h(x,t)$ and (when under pumping) the excitons' common wave function $\psi(x,t)$. The initial wave function $\psi(x,0)$ is normalized to the total number $N$ (or the concentration per site $n$) of pumped excitons:

$$\int_{-L}^{L} dx\, |\psi(x,0)|^2 = N = nL \qquad (1)$$

where $2L$ is the sample length. Henceforth, all lengths $(x, L, |\psi|^{-2})$ are measured in units of the TTF-CA dimer size $d$ which is the lattice period in the N phase. The energy functional density is

$$W(q, h, q_c, \psi) = \frac{a}{2}q^2 + \frac{b}{3}q^3 + \frac{c}{q_c}(q_c - q)h^2 + \frac{f}{2}h^4 + \frac{A}{2d^2}\left(\frac{\partial h}{\partial x}\right)^2$$

$$-gq|\psi|^2 + \frac{k}{2}|\psi|^4 + \frac{\hbar^2}{2md^2}\left|\frac{\partial \psi}{\partial x}\right|^2 \qquad (2)$$

Here $q_c = \rho_c - \rho_n$ is the critical value of $q$ for reaching the instability in $h$ when the energy minimum (in $h$ at given $q$) moves from $h=0$ at $q<q_c$ to $h=\pm h_{eq}$ at $q>q_c$; $g$ is the coupling constant of excitons with the charge transfer $q$, and $k$ is the repulsion energy of excitons. Terms with $x$ derivatives describe inhomogeneous states, $m$ is the exciton's kinetic mass.

*2.2. Physical parameters and estimations*

Now we must relate the constants in eqs. (1) and (4-6) with physical parameters and estimate their values.

$q_I$ : the charge transfer goes from $\rho_N$=0.32 to $\rho_I$=0.52, hence $q_I$=0.2.

**f** : in units of $d$=7.4Å, the dimerization in the I phase is $h_I$=0.03 [16]. Knowing $q_I$ and $h_I$, we get from (3) $f/c=(q_I/q_c-1)/h_I^2$

**b** : we can consider $bq^3$ as an unharmonism with respect to $aq^2$, hence $b \sim a/q_I = 5a$.

**g**: the experimentally known [17-19] downshift $\delta E_{ex}=-gq$ of the exciton energy with increasing $\rho$ fixes $g$=0.4eV.

**k**: repulsion of excitons as dipoles oriented transversely to their distance is estimated as 0.3eV.

**m**: the experimental width of the exciton absorption line gives the estimate $\hbar^2/2md^2$=0.1eV.

**ω, γ** : period of coherent oscillations is T=0.6 ps, then the dimer mode frequency is $\omega=2\pi/T\approx 10^{-2}\text{fs}^{-1}$.

Its relaxation time is [19] $\tau_h$ =3−7ps; taking it as 5ps, the damping parameter is $\gamma=1/\omega\tau_h$ =0.02.

**τ** : For the exciton life time we shall use $\tau=10^{-11}$s, cf. $\tau\sim 10^{-12}$s for singlet excitons in conducting polymers [11]. There are no direct experimental data for these IMEs.

**A**: it is expressed via $\omega$ and the sound velocity as $A/c=s^2/\omega^2\sim 0.02$ from an estimate $s=10^5$cm/s.

**a**: An excursion is necessary to the CTE which energy 0.6eV is known and for which the term $\sim q^2$ in (2) may be considered as an unharmonism in $q/q_N$ which gives $a$=0.1.

**c and $q_c$** : at $T_{NI}$, $W_I=W_N$ and $dW_I/dq$=0 which yields $q_c$=0.0322, c=1.42.

Finally, we are left only with the monitoring parameter $q_c$ dependent on proximity to $T_{NI}$.

## 3. Competition of homogeneous ground states

To illustrate equilibrium properties at a given homogeneous pumping $|\psi|^2=n=cnst$, we consider the functions given by eq. (2) with no $x$ derivatives: $W(q,h,q_c,n))$, and $W_1(q,q_c,n)=W(q,h_{eq}(q),q_c,n)$. Here $h_{eq}(q)$ is given by equilibrium $\partial W/\partial h$=0 over $h$ at a given $q$ which yields the finite dimerization if $q>q_c$:

$$h_{eq}(q) = \sqrt{(1 - q/q_c)c/f} \qquad (3)$$

Plots of $W_1(q,q_c,0)$ in Fig.1 correspond to three regimes separated by two critical levels of $q_c$. The vertical line shows the threshold q= $q_c$ where the dimerization $h$ is turned on, its emergence changes the curvature downwards.

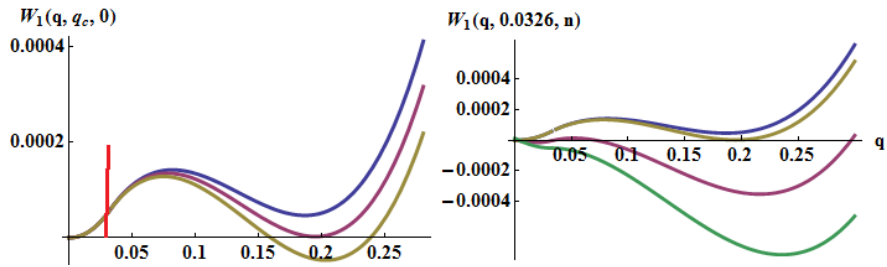

Figure1. Left: $W_1(q,q_c,0)$ as functions of $q$ for different $qc$. From top to bottom: $q_c$=0.0326, $q_c$=0.0322, $q_c$=0.0318. The vertical line is $q=q_c$. Right: $W_1(q,q_c,n)$ as functions of $q$ for a fixed $qc$=0.0326 at different $n$. From top to bottom: $n$=0, n=$n_{c1}$ =0.005, n=$n_d$ = 0.006, n=$n_{c2}$ =0.0095.

At $q_c<q_{c1}$ the I phase appears as a metastable one, at the transition point $T_{NI}$, $q_c=q_{c2}$ the energies of the N and I phases become equal but separated by the barrier. At $q_c<q_{c2}$ the I phase is lower in energy, but the barrier still exist, so the N phase is metastable.

Fig.2 shows 2D contour plots of $W(q,h,q_c,n=0)$ for same values of $q_c$ as in Fig. 1.

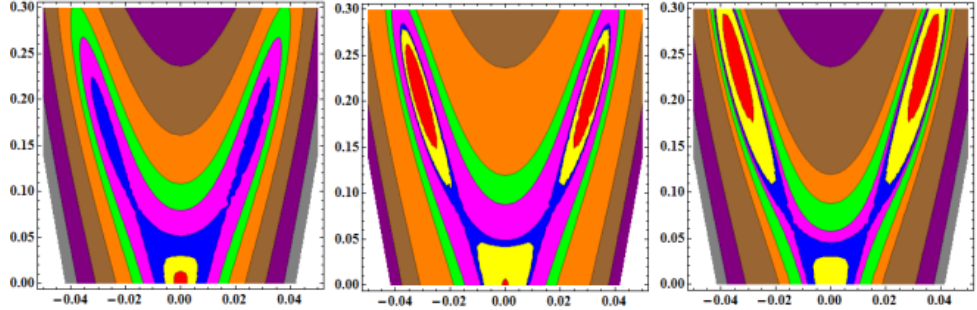

Fig. 2. 2D contour plots for the energy function $W(q,h, q_c,0)$.
Left: $q_c=0.0326$, Centre: $q_c=0.0322$, Right: $q_c=0.0318$

At the left panel we see the stable N state at $q=0$, $h=0$ and two shallow metastable I states at finite $q$ and $h$. At the middle panel both states have the same energy, separated by the barrier. At the right panel, the I state is stable and the N state is metastable.

The pumping of excitons changes the energy landscape. In Fig. 1 (right), we present plots of $W_1(q,q_c,n)$ as a function of q at different pumping levels $|\psi|^2$=n=cnst. The chosen $q_c$=0.0326 corresponds, before pumping, to the stable N and metastable I states but the pumping can change this order. Notice that the pumping enhances the iconicity of the N phase by shifting its equilibrium from initial minimum at $q=0$ to a new $q>0$.

In general, there are three critical levels of the pumping density: two of them $n_{c1}$ and $n_{c2}$ have a thermodynamic origin providing criteria of the 1st and the 2nd order phase transitions, and one - $n_d$ is of a dynamic origin. At $n_{c1}$ the energies of N and I states become equal being separated by the barrier. At $n=n_{c2}$, the barrier disappears, the N phase becomes absolutely unstable. At $n=n_d$, the N state still exists as a metastatble state, the barrier is here, but its height drops down to $W_b$=0 – the level before the illumination so it can be overcome by inertia after the short pulse of pumping (see the general mechanism in [7], and the more specific one in [10]).

## 4. Modelling of Self-trapping evolution

### 4.1. Dynamical equations

We perform the spacio-temporal modeling by considering the joint evolution of the exciton wave function $\Psi(x,t)$, the charge transfer $q(x,t)$, and the lattice deformations $h(x,t)$. We need to solve self-consistently three coupled equations which are generated by variations of the functional (1): the nonlinear Schrödinger eq. for $\Psi(x,t)$, the minimum eq. for $q(x,t)$, and the nonlinear wave eq. for *h(x,t)*.

$$i\hbar \frac{\partial \psi}{\partial t} + i\frac{\hbar}{\tau}\psi = -gq\psi + k|\psi|^2\psi - \frac{\hbar^2}{2m}\frac{\partial^2 \psi}{\partial x^2} \quad (4)$$

$$aq + bq^2 - ch^2 - g|\psi|^2 = 0 \quad (5)$$

$$\frac{1}{2\omega^2}\frac{\partial^2 h}{\partial t^2} + \frac{\gamma}{2\omega}\frac{\partial h}{\partial t} = \frac{s^2}{2d^2\omega^2}\frac{\partial^2 h}{\partial x^2} - \frac{f}{c}h^3 - h(q_c - q) \quad (6)$$

The pumping intensity is introduced via homogeneous initial conditions: *q(x,0)=0* and *h(x,0)=0* while *ψ(x,0)* was almost a constant $\psi(x,0) \approx \sqrt{n}$ with dips at ±*L* to satisfy the box boundary conditions *ψ(±L,t)=0*.

As an illustration, we consider first the enforced homogeneous regime with no x dependences; it can be also viewed as the zero-dimensional regime of an optical dot. Just below the critical *n=0.0945*, *q* is simply displaced and *h* is absent. Just above the threshold there is a sharp dynamic transition followed by oscillations to the state with values of *q* and *h* corresponding to the I phase.

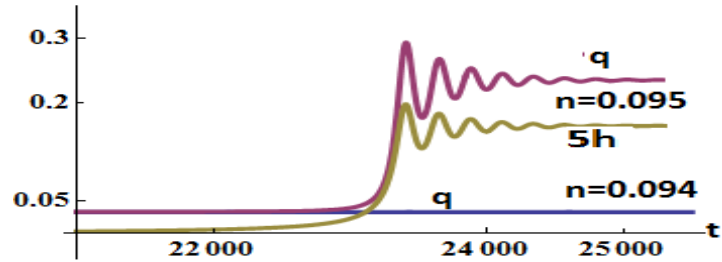

Figure 3. D=0 case. Time dependencies for *q* and *h* at *n* just below and above the critical value *0.0945*.

Fig. 4 shows the space-resolved results for a subcritical pumping $n<n_{c1}$.

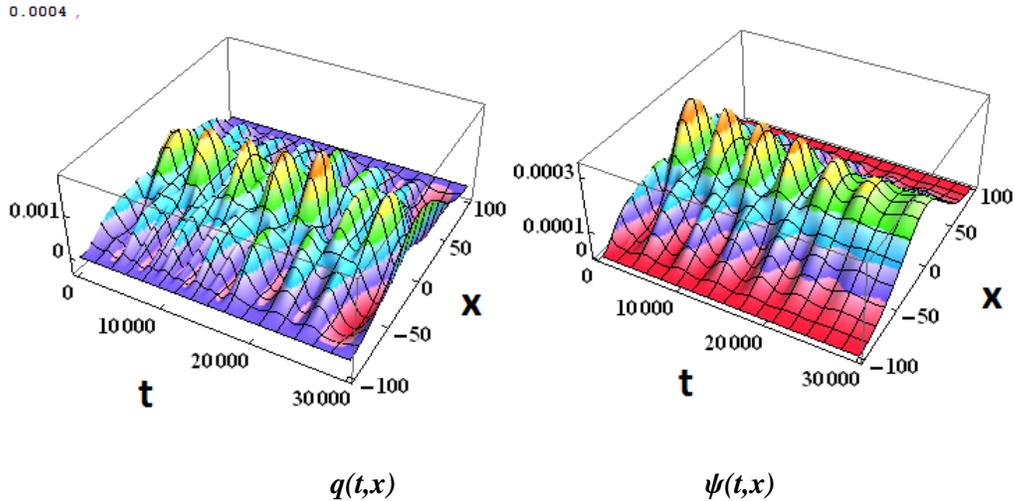

*q(t,x)*           *ψ(t,x)*

Figure 4. Subcritical pumping: n=0.0001, 0<t<30 000fs.

The concentration of excitons is not sufficient to activate $h$ but it provokes some reaction of $q$ which equilibrium (still in the N state) is displaced to q≠0. Both $\psi$ and $q$ get concentrated towards the centre but still are spread over the whole sample. This shape modification gives rise to undulations which are longer seen in $q$ rather than in $\psi$. These are not more regular coherent oscillations with the lattice frequency $\omega$ which are seen in Figs.4,7 – $h$ has not come to the game. There is no more evolution till the life time $\tau$ of the excitons after which everything decays (not shown in Fig.3).

Figs. 5 and 6 show results at higher $n=0.0025$, when the self-trapping develops self-consistently in $\psi$ and $q$, triggering the appearance of $h$ at later times. All together that gives rise to a sharply localized long living domain for all three variables. Thanks to self-trapping effects the critical pumping for the phase transition in the domain is lower that one indicated on Fig.3 for the D=0 case.

Fig.5 shows the evolution in the range of shorter time $t<3000fs$. At $t\approx1000fs$ we see the nucleations of selftrapping in two symmetrical points $x\approx\pm30$. At $t\approx1500$ the two nucleus coalesce into one core which stays since then around $x=0$. The undulations are seen similarly but stronger than for the sub-critical pumping of Fig.4.

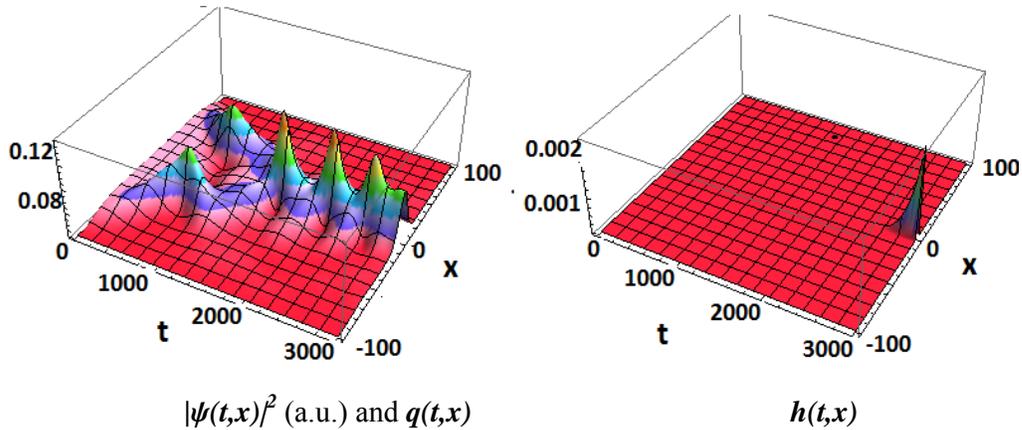

$|\psi(t,x)|^2$ (a.u.) and $q(t,x)$          $h(t,x)$

Figure 5. Supercritical pumping at n=0.0025: Plots for the short time range 0<t<3000fs.

At $t\approx2800fs$ the third event takes place: a sharp raise of $h$ manifests the dynamical symmetry braking phase transition within the same strongly localised domain.

Fig.6 shows the long time evolution (0<t<40000fs) at the same pumping intensity n=0.0025, now taking into account the excitons' damping time $\tau=10^4fs$. Notice that the short domain of $q=0.2$, $h=0.03$, which is just the region of the I state, persist over long time, even at $t>>\tau$ after the excitons have disappeared.

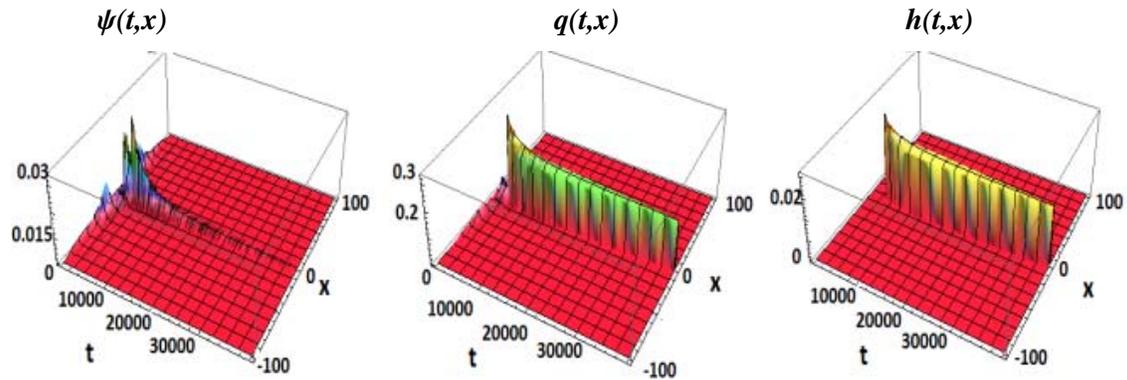

Figure 6. Long time range 0<t<40000fs at n=0.0025, $\tau=10^4$fs.

With a further increase of pumping, the number of domains multiplies at a succession of critical $n_c$; there is no essential change in the domains' width. The case of three domains is shown in Figs.7,8.

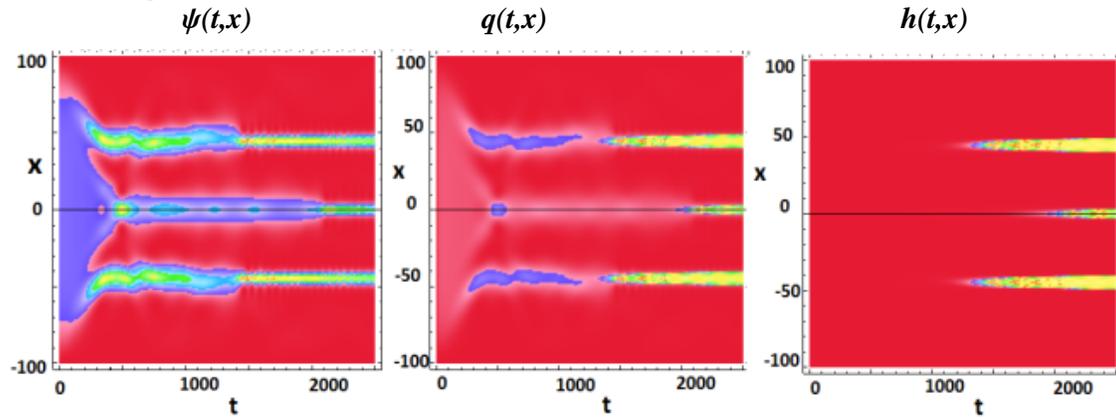

Figure 7. Short time formation and evolution of three I phase domains; n=0.007

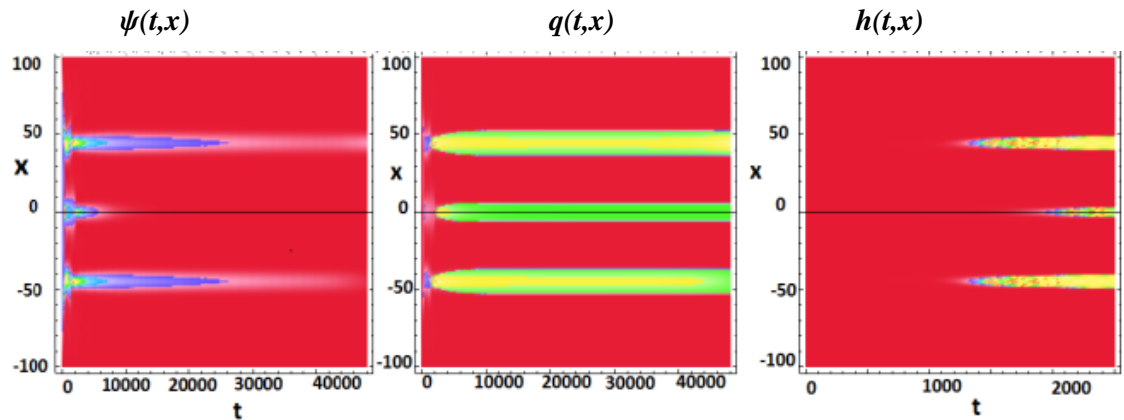

Figure 8. Long time evolution with formation of three I phase domains; $n=0.007$, $\tau=10^4fs$.

As for the case of one I domain in Figs. 5,6, now the lattice deformations *h* also develop much later, via sharp dynamical phase transitions, at times different for domains of different strengths; a typical time is seen in Fig.7 as t≈1500fs.

At larger times the profiles are stabilized and do not evolve anymore till much longer life time τ of the excitons. Coherent oscillations with a frequency ω persist over long times. They are seen in Fig.7 as a fine structure along the time stripes (the yellow background) appearing after emergence of *h*.

## 5. Conclusion

Our primary emphasis was upon the quantum nature of the excitons' motion which forces their delocalization into plain waves. That would happen inevitably for an ideal resonance pumping when the single photon creates the single exciton with the momentum k=0. In reality, the pumping goes with an access energy which gives rise to an exciton in a complex with other modes which total momentum is still zero. The initial relaxation with collisions leaves the exciton as a wave packet rather than a pure state, which is tempting to view as localized at a single molecule or a dimer according to the commonly exploited picture. But the loss of the kinetic energy from such a sharp localization will be ~0.05-0.1eV estimated as a ½ of the exciton's bandwidth. Then with cooling below ~$10^2$K the exciton will descent to the plane wave state at the bottom of its band.

This is where our story begins. The localization will develop as the self-trapping; its length is determined by the balance between gaining the potential energy and loosing the quantum kinetic energy of the exciton. The phase transformation proceeds from large to medium distances, rather than from small to larger ones. The critical concentration of pumping is reached because of the cumulative effect of many delocalized excitons in the quasi-condensate rather than by a local presence of an individual exciton.

The suggested scenario looks to be a minimally sufficient one to take into account the quantum nature of the excitons and their cooperative effects, to understand stabilization of a long string and the efficiency of phase transformations.

The presented first steps call for much further work. Among that, name effects of excitons' repulsion specific to a 1D system like the molecular stack in view. A most important generalization is to study the case of CTE where the internal structure comes to the scenery in an ultimate identification with the CT variable *q*. Here we shall meet closer with concepts of domains as solitons exploited intensively since long time [24,26]. Actually our domains can already be viewed as the pairs NI+IN of solitons glued by the distributed wave function of the excitons. The difference in glue is that it is the centre of mass wave function rather than the internal one of the CTE.

For either case, the kinetics of cooling should be taken into account and the incoherent component of the excitons' ensemble should be added.

There are also requests to experiments which still leave unattended dynamical characteristics of single excitons: their lifetimes, decays to lower excitons and unbound e-h pairs. That is a well known subject of low powers pump-and probe experiments which proved to be so fruitful in studies of optically applied materials, particularly the conjugated polymers, see [11] for references. The EEEL experiments, also proved useful in polymers, would clarify our main suggestion of the appreciable bandwidth of excitons.

As a motivation, we have referred to the best experimentally studied TTF-CA, using its known experimental parameters. We believe in a larger applicability of formulated concepts and their appeal to further optical experiments. Actually any system showing a phase transition at presence of a gapful electronic spectrum should possess some excitons coupled to the order parameter which opens wide research perspectives in PIPTs.

**Acknowledgments**

We acknowledge stimulating discussions with H. Okamoto and D. Mihailovic.